\documentstyle[12pt]{article}
\begin{document}
\begin{titlepage}
\vskip 2cm
\begin{center}

{\Large\bf Real-time tunneling }

\vskip 1cm

{\large
D. A. Demir \footnote{Present Address: ICTP, Trieste, Italy} \\}

\vskip 0.5cm   
{\setlength{\baselineskip}{0.18in}
{Middle East Technical University, Department of Physics, 06531, Ankara, Turkey\\ }} 
\end{center}
\vskip .5cm
\begin{abstract}
Starting with the equivalence of the rest energy of a particle to an
amount of the radiant energy characterized by a frequency, in addition to
the usual relativistic transformation rules leading to the wave-particle
duality, we investigate the case in which this frequency is an internal
propery of the particle. This kind of interpretation of the frequency is
shown to be relevant to the tunneling effect. The investigations in this
direction yield (1) a purely real time everywhere, (2) an anti-hermitian
momentum operator, (3) a corpuscular structure for the particle, and
(4) all of the known theoretical predictions about the tunneling effect.
\end{abstract} 
\end{titlepage}

\section{Introduction}
Quantum mechanical tunneling has become one of the most interesting 
applications of the quantum theory since its early days. It has proven to 
be the key concept in uderstanding the physics of the systems having 
classical stable configurations seperated by an impenetrable energy 
barrier. Tunneling phenomenon is unique to quntum theory and play an 
important role in understanding the various physical processes, ranging 
from birth of the unverse from the nothing \cite{universe} to $\alpha$- 
decay \cite{alpha}.
 
In this work we restrict ourselves to simple quantum systems which can be 
analyzed by the use of the Schroedinger theory. In particular, we are 
not interested in the tunneling phenomena in field theories where one 
faces with an infite number of degrees of freedom (See the recent review 
\cite{japan}). In this sense, analysis of the tunneling process is a text 
book example (See, for example, \cite{merzbach} or \cite{gasio}). In the 
discussion of the stationary tunneling processes where a particle of 
energy $E$ is incident on a potential barrier $V(x)$, entire spatial 
range is classified according to whether $E\geq V(x)$ or $E\leq V(x)$. 
While the former defines the normal region, the latter defines the 
barrier region where the total energy of the particle is below its 
potential energy. In the normal region particle has real-time 
trajectories satisfying the usual Newtonian laws of motion, whereas in 
the barrier region strict energy conservation forces particle to have 
imaginary-time trajectories \cite{mclau,carlitz,hols}. As is 
well-known, in the barrier region Schroedinger equation has evanescent- 
wave solutions characterized by decaying and growing exponentials. 
Despite the well-understood formalism of the tunneling summarized here, 
the debate about the tunneling time continues to exist. This mainly 
stems from the fact that time is a parameter in the quantum theory, and 
consequently, one does not have an agreed-upon formalism to calculate it 
\cite{hau-stov}.
  
It would be convenient to use every oppurtunity to have a better 
insight into the tunnelling phenomenon as it might be useful in 
arriving at the full solution (including the tunneling time itself) of the 
problem. In this work our aim is not to calculate the tunneling time. We 
shall just present a different way of analyzing the tunneling effect to 
gain more insight into its nature. The method of the analysis can be 
summarized by mentioning that we introduce different Hamiltonians for 
normal and barrier regions; in particular particle is endowed with a 
negative kinetic energy in the barrier region. As will be seen in the 
text negative kinetic energy is not an ad hoc assumption, instead, it 
will be shown to follow from the first principles. The main results of 
the work can be summarized as  the reality of the time everywhere, 
non-hermitian character of the momentum in the barrier region, 
corpuscular nature of the particle in the barrier, and the reproduction of 
all of the known results.

In Sec.2 we mention the formalism and discuss its phenomenological 
viability in the classsical and quantum limits. 

In Sec.3 we present a detailed discussion of the results and compare them 
with those of the usual formalism.

\section{Derivations}
Let us consider a relativistic particle of rest mass $m_{0}$. As 
usual, one can equate its rest energy $m_{0}c^2$ to a unit of Planckian 
energy \cite{debrog,barut} 
\begin{eqnarray} 
m_{0}c^2=\hbar\omega_{0}
\end{eqnarray}
where the emphasis is on the quantum character of this equation as it looses
its meaning as $\hbar\rightarrow 0$. In what follows we require (1) to be
covariant under Lorentz transformations, that is, it will hold in an
arbirary inertial frame $K$ moving with velocity $v$ relative to the rest
frame $K_{0}$ in which (1) holds.

To determine the form of (1) in the moving frame $K$ one should know the
transformation rule for the frequency and mass. Actually, knowing the
transformation rule for either of them is sufficient as long as the 
equality (1) holds in any inertial frame. Here we shall 
base discussion on the transformation properties of the frequency, and
conclude that of the mass from the covariance. There are two
known options one can follow in specifying the transformation rule
for the frequency \cite{landau}. First, in $K_{0}$ there can be a plane
wave $e^{i\omega_{0}t_{0}}$ accompaniying or co-existing with the
particle. As the phase of a plane wave is a relativistic invariant, in the
moving frame $K$ equation (1) becomes a four-vector relation:
\begin{eqnarray}
P_{\mu}=\hbar k_{\mu}
\end{eqnarray}
which is nothing but the usual relation between energy-momentum 
four-vector $P_{\mu}=(m c^2, c\vec{P})$ and the wave-vector 
$k_{\mu}=(\omega, \vec{k})$. It is in this sense that the particle is 
associated with a wave motion having the propagation vector $k_{\mu}$. 
The expression for $m$ is the usual one
\begin{eqnarray}
 m &=&\frac{m_{0}}{\sqrt{1-v^2/c^2}}
\end{eqnarray}
where $m_{0}$ is related to $\omega_{0}$ via (1). The de Broglie relation 
(2) is the basic statement of the wave-particle duality allowing for the 
representation of the quantum mechanical particles by the wave phenomena, 
leading to wave mechanics.

Until now we have rederived the known statement of the wave-particle 
duality (2) from (1) by making the standard assumption that the particle is 
accompanied by a wave propagation. However, this transformation property  
of the frequency is just one alternative; that is, $\omega_{0}$ in (1) can 
well be an internal property of the particle itself, in which case  
it is viewed as a clock of frequency $\omega_{0}$ in $K_{0}$. After 
ascribing this meaning to $\omega_{0}$, in the moving frame $K$ equation 
(1) takes the form 
\begin{eqnarray}
M=\hbar\Omega\; .
\end{eqnarray} 
Being one of the most fundamental results of the theory of 
relativity moving clocks slow down, that is 
$\Omega=\omega_{0}\sqrt{1-v^2/c^2}$, so the moving frame mass $M$ is 
related to its rest frame value via 
\begin{eqnarray}
M=m_{0}\sqrt{1-v^2/c^2}.
\end{eqnarray} 
Now let us remark on some consequences of this transformation rule.
Equation (4), as opposed to (2), is not a four-vector relation, 
instead it just reexpresses (1) in the moving frame $K$, and there is no 
propagation characteristics ( like $\vec{k}$ in (2)) whatsoever in it. 
Hence, it must be emphasized that with this interpretation of $\omega_0$ 
in (1) particle could be viewed as a corpuscule having no wave-like 
properties. Here, by corpuscule, we mean a localized object where the 
degree of localization varies from a point (classical material point) to 
a distribution (quantum distribution). 

Furthermore, the transformation rule for mass in (5), unlike the one in (3),
is completely unrealistic as it contradicts with the well-established 
experimental facts. In fact, it was merely because of this 
reason that de Broglie eliminated this transformation rule.
Fortunately, however, the strange transformation rule (5) will be seen 
to match the requirements of the quantum tunneling.
 
Taking the non-relativistic limit of equations (3) and (5), using the 
usual formula for kinetic energy $(m-m_{0})c^2$, and adding an arbitrary 
potential function $V(x)$ to take into account the possible interaction 
of the particle with the environment, one gets the following energy 
functions 
\begin{eqnarray}
(3)\leadsto h&=&\frac{1}{2}m_{0}(\frac{d\,x}{d\, t})^{2}+V(x)\\
(5)\leadsto H&=&-\frac{1}{2}m_{0}(\frac{d\,x}{d\, t})^{2}+V(x)
\end{eqnarray}
We note that while kinetic energy of the particle is positive-definite 
in the wave representation (6), it is negative-definite in the case of 
corpuscular representation (7). Furthermore, as the basic equality (1) 
guarantees, the energy functions (6) and (7)  become identical if the 
velocity vanishes in both representation. We now analyze (6) and (7)
in detail to investigate their physical implications.
\subsection{Classical Considerations}
Let us start discussing (6). For a time-independent potential, $V(x)$, 
energy is conserved, and $h=E$ is a first integral of the motion. Therefore
the trajectory of the particle can be shown to satisfy 
\begin{eqnarray}
m_{0}\frac{d x}{d t} &=& \sqrt{2m_{0}(E-V(x))}\\
m_{0}\frac{d^2 x}{d t^2} &=& -\frac{d V}{d x}\; .
\end{eqnarray}
The first equation requires $E\geq V(x)$ for the particle to have a real 
momentum. The next one is the well-known statement of the Newtonian laws 
of motion. As a result, the energy function (6) leads us to the 
well-known description of the classical systems for which the total 
energy of the system exceeds or at most equals the potential energy field 
on it, as is always the case.
  
Repeating the same steps of analysis for (7), with $H={\cal{E}}$ is a 
constant of the motion, one obtains
\begin{eqnarray}
m_{0}\frac{d x}{d t} &=& \sqrt{2m_{0}(V(x)-{\cal{E}})}\\
m_{0}\frac{d^2 x}{d t^2} &=& \frac{d V}{d x}\; 
\end{eqnarray}  
which have entirely different implications on the particle dynamics 
than (8) and (9). The first equation requires ${\cal{E}}\leq V(x)$, for 
particle's momentum be real. The next equation, on the other hand, forces 
practically particle to obey Newtonian laws for an inverted potential 
$V(x)\rightarrow -V(x)$, that is, one does not have the true laws of 
motion for the trajectory of the particle. Consequently, the energy 
function $H$ in (7) restricts particle to those regions of the space in 
which the total enery of the particle is at most equal to the
potential energy on it. 

If we require, for both $h$ (6) and $H$ (7), particle be endowed with 
real position and time coordinates we should restrict these energy 
functions to their appropriate regions of validity, that is, we should use 
$h$ when $E\geq V(x)$, and $H$ when $E\leq V(x)$ for a particle with total 
energy $E$. With this prescription in mind, one would recall the 
tunneling effect where such distinctions as $E\geq V(x)$ and $E\leq 
V(x)$ are of fundamental importance for the analysis of the problem.
In fact, all these unusual properties (10) and (11) have have already    
been observed and utilized in the analysis of the quantum mechanical   
tunneling by path integral methods \cite{mclau,hols}. One should note,   
however, there is a fundamental distinction between the two; namely, (10)
and (11) have been derived without introducing an Eucledean time, unlike
the path integral formulation of the quantum tunneling where rotation of
the time parameter from real to purely imaginary values is indispansable.

To combine the implications of wave- and corpuscular- representations 
we have analyzed till now, it would be convenient to refer a typical 
potential energy graph, like the one shown in Fig.1. In drawing Fig.1 we 
assumed a particle with coordinate $x$ and total energy $E$. The points 
labelled by $a$, $b$ and $c$ are the classical turning points in which 
$V(x)=E$. In the discussion below we assume both energies $E$ in (8) and 
${\cal{E}}$ in (10) are identical and equal to $E$, as suggested by Fig. 1.

For $a\leq x \leq b$ we have $E\geq V(x)$, so we conclude that in this 
region the total energy of the system has the expression in (6), and 
dynamical evolution of the particle's coordinate is governed by (9). 
The boxed label $h$ stands for the energy function $h$ in (6). While the 
velocity of the particle (8) vanishes as it approaches to the  
turning points $a$ or $b$, it is accelerated through the equation of 
motion (9) back to the region of incidence. In this way, under the 
strict energy conservation, particle moves back and forth between the 
turning points $a$ and $b$. The half-period, $T$, of the motion can be 
found \cite{golds} by inverting (8) for time
\begin{eqnarray}
T_{h}=m_{0}\int_{a}^{b} \frac{d x}{\sqrt{2 m_{0} (E-V(x))}} \, .
\end{eqnarray} 

For $b\leq x \leq c$, however,  we have $E\leq V(x)$ so we conclude 
that here the total energy of the system has to have the expression in 
(7), and dynamical evolution of the particle's coordinate is to be 
governed by (11). The boxed label $H$ stands for the energy function $H$ 
in (7). While the velocity of the particle (10) vanishes as it 
approaches to the turning points $b$ or $c$, it is accelerated through 
the equation of motion (11) back to the region of incidence. In this way, 
under the strict energy conservation, particle moves back and forth 
between the turning points $b$ and $c$. The half-period, $T_{H}$, of the 
motion can be found by inverting (10) for time
\begin{eqnarray}
T_{H}=m_{0}\int_{b}^{c} \frac{d x}{\sqrt{2 m_{0} (V(x)-E)}} \, .
\end{eqnarray}

We conclude from these discussions that particle oscillates indefinitely 
whichever region it happens to fall initially. It remains trapped in 
the associated region unless the configuration of potential energy 
is modified by some external agent. From the realistic point of view, the 
$h$-region $a\leq x \leq b$ is the one where classical systems can exist, 
as is evidenced by the appearence of the usual equation of motion (9).
The trajectory of the particle in $H$-region, $b\leq x \leq c$ is no way 
classical as dictated by its equation of motion (11). Although form of 
(11) is exactly the one one would use in analyzing the  tunneling effect 
by path integral methods, its derivation does not rest on a Wick 
rotation of the time parameter; it follows directly from the energy 
function $H$ (7). Hence both half-periods $T_{h}$ and $T_{H}$ are 
intrinsically real and represent the oscillatory character of 
the motion in the associated regions. 
  
As usual, one would call the transition of the particle from the left 
($x\leq b$) of the barrier to the right ($x\geq c$) as tunneling. The 
classical considerations we have followed till now do not provide a means 
for particle to transit from $h$-region to $H$-region and vice versa. For 
tunneling to occur, as is phenomenologically the case, there should be 
a cause for particle be hopping from the $h$-region trajectory to the 
$H$-region trajectory, which we necessarily attribute to quantum effects 
to be discussed below. 
\subsection{Quantum Considerations}
In this section we shall discuss the issue of quantization to understand 
the tunneling effect. Let us first consider a particle in $h$-region whose 
position and momentum at time $t$ are represented by operators $\hat{x}(t)$ 
and $\hat{p}(t)$, respectively. These operators do satisfy the fundamental 
quantization postulate
\begin{eqnarray}
[\hat{x}(t),\hat{p}(t)]=i\hbar\, .
\end{eqnarray}
In the position basis, for example, the representation $\hat{x} = x$ and 
$\hat{p} = -i\hbar \frac{\partial}{\partial x}$ satisfies the quantum 
bracket (14). The Hamiltonian operator $\hat{H}=\hat{p}^2/2m_{0} + 
V(\hat{x})$, after making the position basis replacements for 
$\hat{x}(t)$ and $\hat{p}(t)$, leads us to the Schroedinger equation 
for $\psi(x,t)$
\begin{eqnarray}
\{ -\frac{\hbar^2}{2m_{0}}\frac{\partial ^2}{\partial x^2} + V(x) \} 
\psi(x,t) =i\hbar\frac{\partial}{\partial t}\psi(x,t) \; .
\end{eqnarray}
The stationary state solution (as is natural for a time-independent 
$V(x)$) for a slowly varying potential is the well known WKB wavefunction 
\begin{eqnarray}
\psi(x,t)\sim \frac{1}{\sqrt{p(x)}}
\exp{\frac{i}{\hbar}(-Et+\int^{x} dx'p(x'))}
\end{eqnarray}
where $p(x)=\sqrt{2m_{0}(E-V(x))}$. As expected wavefuction is of 
oscillatory character, and diverges at the turning point as a by-product 
of the WKB approximation.

Now let us dicuss the issue of quantization for the corpuscular  
representation for which the main object of the discussion is the energy 
function $H$ (7). First of all, $H$ leads to the Hamiltonian function 
$-p^2/2m_{0}+V(x)$ where $p=-m_{0}\frac{d x}{d t}$. If we quantize this 
Hamiltonian with the usual quantization prescription (14), we  
necessarily obtain a wrong sign kinetic energy operator 
$\frac{\hbar^2}{2m_{0}}\frac{\partial ^2}{\partial x^2}$ which, when 
substituted in the Schroedinger equation (15), leads to an oscillatory 
wavefucntion like (16), except for $p(x)\rightarrow \sqrt{2m_{0}(V(x)-E)}$. 
Such a solution is obviously unphysical because transition probability  
is expected to have an exponential fall-off with the barrier width as has 
already been confirmed by the experiment in various circumtances. 
Therefore, to overcome this difficulty one should find a way out, namely, 
one should apply an appropriate quantization procedure in treating the 
particle in corpuscular representation. In searching for the appropriate 
quantization rule, one notices that Hamiltonian function for the 
corpuscular representation would have the same form as that of the wave 
representation if the momentum is allowed to take imaginary values 
$p\rightarrow i p$. Following the clue provided by this observation, the 
position and the momentum of the particle in the corpuscular 
representation are represented, respectively,  by the operators $\hat{x}(t)$ 
and $\hat{p}(t)$ subject to the quantization rule 
\begin{eqnarray}
[\hat{x}(t),\hat{p}(t)]=\hbar\, .
\end{eqnarray}
Then, in the position basis the representation $\hat{x} = x$ and
$\hat{p} = -\hbar \frac{\partial}{\partial x}$ satisfies this quantization 
condition. Obviously, momentum operator is anti-hermitian and its 
consequences will be discussed below. Using this position basis 
representation we obtain the Schroedinger equation for $\psi(x,t)$
\begin{eqnarray}
\{ -\frac{\hbar^2}{2m_{0}}\frac{\partial ^2}{\partial x^2} + V(x) \}
\psi(x,t) =i\hbar\frac{\partial}{\partial t}\psi(x,t) \; .
\end{eqnarray}
whose stationary state solution for a slowly varying potential is the 
well- known under-barrier WKB wavefunction 
\begin{eqnarray}
\psi(x,t)\sim \frac{1}{\sqrt{\tilde{p}(x)}}
\exp{\frac{i}{\hbar}(-Et+i\int^{x} dx'\tilde{p}(x'))}
\end{eqnarray}
where $\tilde{p}(x)=\sqrt{2m_{0}(V(x)-E)}$. It is obvious that this 
wavefunction has non-oscillatory character in space and has the same form 
as one would obtain from the usual quantization procedure. This completes 
the quantization issue of wave- and corpuscule-like representations of 
the matter which have direct relevance to the tunneling phenomenon.
Now we shall turn to a comparative discussion of the present formalism 
and the usual one to point out the extra insight brought by the 
corpuscular representation.
\section{Discussions and Conclusions}
We now compare the results of this work with those of the usual formalism 
in a comparative manner. We first summarize the discussion of the 
tunneling phenomenon in usual terms and touch on some important points 
both in classical and quantum mechanical regimes.
\begin{enumerate}
\item In the usual formalism one starts, at the classical level, with 
the energy function $\frac{1}{2}m_{0}(\frac{d\,x}{d\, t})^{2}+V(x)$ which is 
valid everywhere provided that the associated Lagrange function yields the 
correct Newtonian laws of motion. As the Hamiltonian does not 
have an explicit time dependence it is conserved, and the total energy $E$ 
is a constant of motion. As discussed in detail in Sec. 2.1, in this case 
particle experiences an oscillatory motion with a period twice (12). In 
those portions of space where $V(x)\geq E$, momentum of the particle 
$\sqrt{2m_{0}(E-V(x))}$ turns to imaginary and for consistency one needs 
to make $m_{0}\frac{d x} {d t}$ imaginary. As a natural choice, one 
continues time to pure imaginary values: $t\rightarrow -it$. Therefore, 
at the classical level of the discussion tunneling event could be 
concluded to cost no real time at all. 
\item In the above-mentioned classical setting, one applies the quantization 
rule (14) and derives the associated wave equation for $\psi(x,t)$ which 
is valid everwhere in space-time. In doing this one assumes that 
-referring the discussion at the beginning of Sec.2- the rest energy of a 
(masive) particle is equivalent to an amount of the Planckian energy 
which is always associated with a wave propagation- as we named in the 
text 'wave representation'.  The Schrodinger equation, which is a 
differential equation that that wave motion is to 
satisfy in the non-relativistic interacting limit, describes the 
dynamics of the particle everywhere. In the tunneling region, where 
$V(x)\geq E$, it produces evanescent waves which exponentially decay (or 
grow) in space. A similar situation occurs in solving the 
Maxwell equations, for example in conducting media, and is a sign of the 
power loss in the medium. In such cases one does not have a wave 
propagation, instead a localized distribution showing the fact that (in 
the case of quantum mechanics) it will be less and less probable to find 
the particle in a certain portion of space if one goes away and away from 
the point of entrance into the barrier region. 

Now we turn to the discussion of the tunneling time \cite{hau-stov,others}. 
As is well known, time is not representable by an operator in quantum 
theory, and there is no unique way of calculating how long it will take 
for the particle to appear at the opposite side of the barrier. We have 
discussed nature  of the classical time for the barrier region in Item 1: 
independent of the parameters of the particular problem it is purely 
imaginary; saying that tunneling occurs in zero real time, modulo the 
will-be quantum mechanical contribution.
\end{enumerate}
Now we start discussing the implications of this work about the mechanism 
of the tunneling phenomenon. According to the present work we deal with 
two sets of relativistic transformation laws. The first group applies 
all the observable systems and have already been confirmed 
phenomenologically. In this case one naturally arrives at the well-known 
equations describing the Lorentz transformations, non-relativistic 
equations of motion, and wave-particle duality. We called this type of 
interpretation as 'wave representation' in the text.
    
The second group applies no observed system and includes strange laws, for 
example, energy of a system decreases with its speed and vanishes
when it reaches the speed of light. In this case, the frequency in 
the Plackian energy is an internal property of the system itself, and one 
does not have a wave propagation coexisting with the particle at all. In 
fact, the system at rest is a clock tickling with a frequency proportional 
to its rest mass (1). Transformation to a moving frame just rescales the 
rest-frame parameters without leading one to a four-vector relation as in 
the wave representation. We named this representation as the 'corpuscular 
representation' in the text.  

Needless to say, two transformation laws become identical in the rest frame. 
This work proposes that the non-relativistic interacting Hamiltonian for a 
(massive) particle should be written as in the first group of laws 
when it is in normal regions where Newton's equation of motion is valid, 
and as in the second group when it is not. 

The quantum tunneling provides a unique chance where the above 
classification may find a place of application. This can be understood by 
the following observations: Particles outside the barrier region are under 
direct observation and have phenomenologically well- established 
properties, so that here one can apply the laws of motion residing in 
the first group. Next, at the classical turning points particle is 
momentarily at rest for which one has the situation described in (1). 
Finally, away from the turning point, particle may either return back to the 
normal region or enter the barrier region in which about the dynamics 
of the particle one does not have direct information. Hence, proposing that
particle in the barrier region can be described by the second group of laws 
- as long as consistent - is not forbidden a priori. 
\begin{enumerate}
\item Unlike the usual case which ascribes a single Hamiltonian function 
following from (6), in accordance with these arguments we assume equations 
(6) and (7) to describe the dynamics of the particle outside and inside 
the barrier, respectively. As discussed in detail in Sec. 2.1, under 
such a prescription one arrives at classical trajectories resting exclusively
on real-time trajectories in both normal (8) and barrier (10) regions. 
Depsite the purely real nature of the time parameter, one obtaines the 
same classical equations of motion (see (9) and (11)) appearing in the usual 
discussion of the problem. Thus at the classical level of the discussion 
one concludes that time remains real for both normal (as it should) and 
barrier regions.
\item In quantizing this classical setting, one adopts the quantization 
prescriptions (14) and (17) for Hamiltonians in normal and barrier 
regions, respectively. Like the Hamiltonian functions themselves, the 
quantization prescriptions do also vary as one changes from normal to 
barrier regions. These quantizations rules do naturally lead us to the  
Schroedinger equation appropriate for the region under consideration (see 
(15) and (18)). 

Unlike the 'evanescent wave' characterization of the usual discussion, the 
exponentially decaying character of the under-barrier wavefunction can now
be attributed to the corpuscular nature of the particle consistent with 
the discussions in the previous sections. One can envisage the appearence 
of the 'corpuscular' nature of the particle as its effective localization 
over one de Broglie wavelength. That one arrives at the same wavefunction 
(19) as one would obtain by the known methods is a mathematical  
statement, physically (19) can be interpreted as the quantum 
generalization  of the classical material point. This can be seen by 
observing that as $\hbar\rightarrow 0$, $\psi(x,t)$ in (19) vanishes 
indicating the fact that particle returns back to its classical 
representation of vanishing extension (a material point) and is pushed 
outside the barrier region; the only place ($h$-region) it is allowed to 
exist classically. Thus the interpretation of the frequency $\omega_0$ in 
(1) as some internal property of the particle itself leads one eventually 
to a quantum generalization of the classical material point. On the basis 
of these observations one would understand the mechanism of the tunneling 
phenomenon as occuring when the effective spatial extension of the particle 
becomes comparable to the barrier width.
 
We now discuss nature of the momentum operator in two approaches. As 
tunneling occurs always with $V(x)\geq E$, it is meaningless to consider 
states with definite momentum, except for the idealized case of a 
semi-infinite flat potential for which the Schrodinger equation (18) has 
the stationary -state solution $\phi(x)= constant\times 
\exp{-\frac{p x}{\hbar}}$, where $p=\sqrt{2m_{0}(V_{0}-E)}$, and $V_0$ is 
the barrier hight. In the standard analysis of the tunneling effect 
momentum operator $-i\hbar \frac{\partial} {\partial x}$ is hermitian, and 
$\phi(x)$ is an its eigenfunction with the eigenvalue $i p$. On the 
other hand, in the present approach, momentum operator $-\hbar 
\frac{\partial} {\partial x}$ is not hermitian, and $\phi(x)$ is an its 
eigenfunction with the eigenvalue $p$. Thus, we conclude that in the 
present approach momentum operator reproduces the decaying and growing 
exponentials with real eigenvalues, whereas in the usual approach one 
should introduce purely imaginary eigenvalues to obtain the same 
eigenfunctions. 

Finally we comment on the tunneling time. The amount of time a particle 
spends to traverse a given potential barrier cannot be evaluated in the 
formalism of this work. Then, it remains to apply to certain 
phenomenological methods developed already \cite{others}.
\end{enumerate} 
\section{Acknowledgements}
It is a pleasure for author to express his gratitude to A. O. Barut 
for his highly useful remarks and suggestions during the Conference
on Frontiers in Theoretical Physics at International Center for
Physics and Applied Mathematics, Edirne, Turkey, 1994, where part of this
work was presented. Author thanks to M. Durgut for his helpful 
comments. Thanks also go to N. Akman and {\c S}. Demir for discussions.

\newpage
\begin{figure}
\vspace{7.0cm}
\end{figure}
\begin{figure}
\vspace{12.0cm}
    \includegraphics{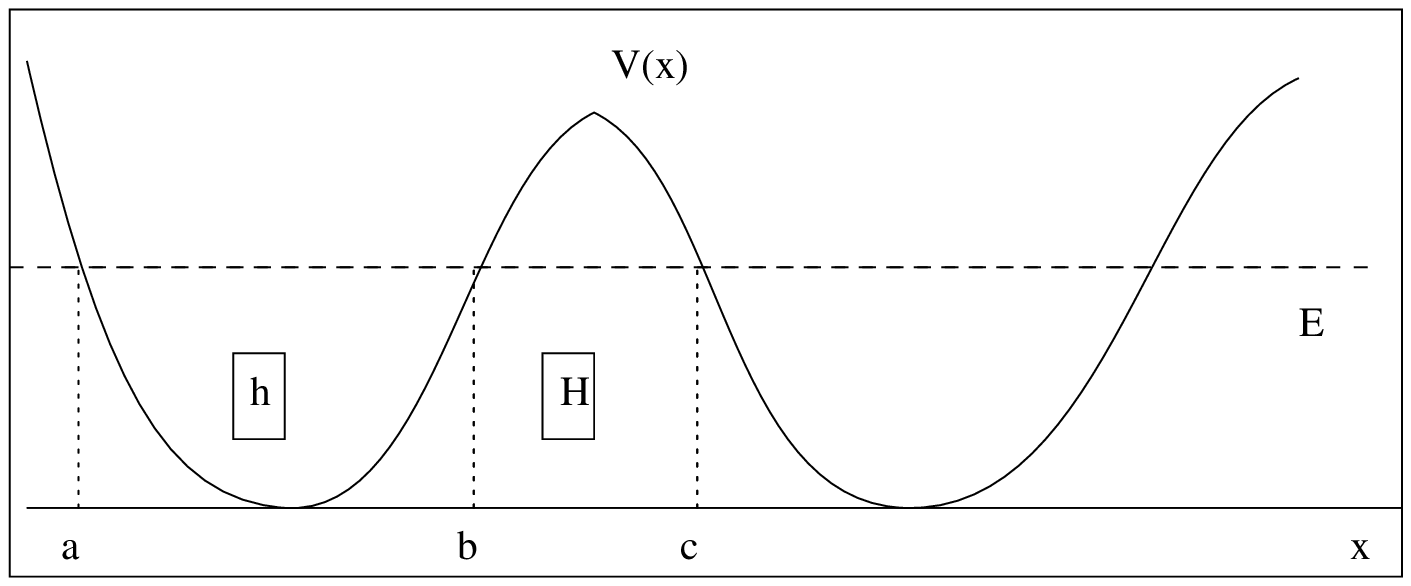}
    \vspace{-8.0cm}
\vspace{0.0cm}
\mbox{ \hspace{0.2cm} \large{\bf Figure 1: A typical potential realizing  
the quantum}}
\mbox{ \hspace{0.2cm} \large{ \bf mechanical tunneling
(see text).}}
\end{figure}  
\end{document}